\documentclass[twocolumn,preprintnumbers,showpacs,amsmath,floatfix,prb,aps,superscriptaddress]{revtex4}

\usepackage{graphicx}
\makeatletter

\begin{document}

\newcommand{\be}{\begin{equation}}
\newcommand{\ee}{  \end{equation}}
\newcommand{\ba}{\begin{eqnarray}}
\newcommand{\ea}{  \end{eqnarray}}

\title{Four-terminal resistance of an interacting quantum wire with weakly invasive contacts}

\author{Hugo Aita} \affiliation{Departamento de F\a'{i}sica and IFLP,
  FCE, Universidad Nacional de La Plata, cc 67 (1900) La Plata,  Argentina}

\author{Liliana Arrachea} \affiliation{Departamento de F\a'{i}sica,
  FCEyN and IFIBA, Universidad de Buenos Aires, Pabell\'on 1, Ciudad
  Universitaria, 1428 Buenos Aires, Argentina}

\author{Carlos Na\'on} \affiliation{Departamento de F\a'{i}sica and IFLP,
  FCE, Universidad Nacional de La Plata, cc 67 (1900) La Plata,  Argentina}

\date{\today}

\begin{abstract}
We analyze the behavior of the four-terminal resistance, relative to the two-terminal resistance
 of an interacting quantum wire
with an impurity, taking into account the invasiveness of the
voltage probes. We  consider a one-dimensional Luttinger model of
spinless fermions for the wire. We treat the coupling to the voltage
probes perturbatively, within the framework of non-equilibrium Green
function techniques. Our investigation unveils the combined effect
of impurities, electron-electron interactions and invasiveness of
the probes on the possible occurrence of negative resistance.
\end{abstract}
 \maketitle


\section{Introduction}
\label{} Quantum transport in novel materials, is one of the most
active areas of present research in condensed matter physics
\cite{QT}. The problems that arise are specially interesting in
one-dimensional (1D) devices such as quantum wires and carbon
nanotubes. In these cases the effect of electron-electron (e-e)
interactions is crucial, leading to the so called Luttinger liquid
(LL) behavior \cite{1D}, characterized by correlation functions
which decay with interaction-dependent exponents
\cite{Exp-exponents} and a power law in the tunneling $I-V$
characteristic curve \cite{powlawcond}.

The actual nature
of the resistance in a mesoscopic device has been a central issue
since the first milestones in the theory of quantum transport. Landauer proposed
 the famous setup to study quantum transport where a mesoscopic sample is placed
between two reservoirs at different
chemical potentials. \cite{cond} Then,
  B\"uttiker \cite{condbut}, in agreement with experiments \cite{exp-cond}, showed
the fundamental relation $G= n G_0$ for the two-terminal conductance
of a non-interacting quantum wire, being $n$ the number of
transverse channels and  $G_0=e^2/h$ the universal conductance
quantum. The remarkable consequence of this simple law is the fact
that a purely non-interacting electronic system without any kind of
inelastic scattering mechanism has a sizable resistance, which for a
single channel device is as large as $G_0^{-1} \simeq 13 k\Omega$.
This resistive behavior is due to the coupling between the system
and the reservoirs through which the driving voltage is applied. For
this reason, this quantity is identified as the {\em contact
resistance} of the ideal non-interacting setup. The mesoscopic
community became then motivated towards the definition of an
alternative physical concept to describe the resistive behavior of
the sample, free from the effects of the contact resistance. In
another pioneering work \cite{engqan}, a {\em gedanken} setup was
proposed in order to sense the local voltage and the temperature.
The main idea is to consider the mesoscopic system locally coupled
to voltage probes or thermometers, represented by means of particle
reservoirs. The latter have chemical potentials or temperatures that
satisfy the condition of local electrochemical or thermal
equilibrium with the mesoscopic system, which implies that the
chemical potentials and temperatures of these systems are adjusted
 in order to get a vanishing
electronic and heat flows through the contacts  to the central
device. For the case of two voltage probes connected along the
sample as in the sketch of Fig. \ref{fig1}, the voltage drop
corresponding to the chemical potential difference
$(\mu_1-\mu_2)/e$,  defines the {\em four terminal resistance} \be
\label{r4t} R_{4t}= \frac{ \mu_1- \mu_2}{e I}, \ee where $I$ is the
current flowing through the setup.
 This scheme to define the four terminal resistance was later
implemented in the framework of scattering matrix theory for
multiterminal setups \cite{but4pt} in wires of non-interacting
electrons with a single \cite{but89,grambu} and many impurities.
\cite{gop} In Refs. \onlinecite{but89,grambu}  it is clarified that
the inference of $R_{4t}$ from a calculation based in a two-terminal
geometry and the original Landauer formula\cite{cond} may not always
be correct, which stresses the importance of considering a genuine
four-terminal setup to properly evaluate this quantity.

 Among other interesting features, for non-interacting systems, it was predicted
that negative four terminal resistances are possible.
\cite{but4pt,but89,grambu,gop,pern} This is a consequence of the
coherent nature of the electronic propagation along a sample where
only elastic scattering processes with barriers or impurities can
take place. These negative (longitudinal) resistances in ballistic
structures were first measured in the late eighties
\cite{takagaki-etal}. More recently, this effect was also observed
in semiconducting structures \cite{pic}. A bit later, the behavior
of $R_{4t}$ was experimentally studied in carbon nanotubes
\cite{gao}. In this case, a negative value of this resistance was
also observed within the low temperature regime. Let us mention that
the four-terminal resistance, in the context of uncorrelated
fermions, has been also extended to the case of time-dependent
voltage probes, leading to the concept of four-terminal impedance
\cite{fed}.
 It is widely accepted that the Luttinger model of interacting electrons in 1D
 is able to capture the main features observed in the transport
experiments of carbon nanotubes \cite{lutran,dolc}. In particular, the
power law behavior of the tunneling current as a function of the
applied voltage and/or temperature predicted from Luttinger liquid
theory has been experimentally observed in these systems. Regarding
the behavior of $R_{4t}$ evaluated from a multiterminal setup in
Luttinger liquids, the literature is restricted to Ref.
\onlinecite{ANS}.
 Previous estimates for this quantity were done on the basis
of an interpretation of Landauer formula in a two terminal setup.
\cite{egger} This is due to the fact that quantum transport in
multiterminals Luttinger liquids or models of interacting electrons
is, in general, a rather challenging problem from the technical
point of view. Besides Ref. \onlinecite{ANS}, genuine multiterminal
systems have been  considered in Y-type geometries
\cite{chamon,lao},  within linear response in the voltage and
Hartree-Fock approximation of the interaction, respectively as well
as in the study of the tunneling current of a quantum wire in the
Fabry-Perot regime. \cite{pug} There are also some recent works on
the effect of wires that are capacitively coupled to an additional
reservoir. \cite{caz,zor}

In Ref. \onlinecite{ANS} we have considered the setup of Fig.
\ref{fig1}, where an infinite Luttinger wire with a single impurity,
through which a current $I$  flows as a response to an applied
voltage $V$, is connected at two points to voltage probes. Following
the  procedure  of previous works for  non-interacting electrons, we
have considered \cite{engqan,but89,grambu,gop,pern} {\em
non-invasive} contacts between the wire and the voltage probes. We
have shown that the voltage profile displays Friedel-like
oscillations, as in the case of non-interacting electrons
\cite{but89,pern,grambu}, but modulated by an envelope displaying a
power law behavior as a function of the applied voltage or
temperature, with an exponent depending on the electron-electron
interaction strength. However, it is known that in the opposite
limit of strong enough coupling between the mesoscopic device and
the probes,
 inelastic scattering events and classical resistive behavior take place
 \cite{dampast}.
Moreover, ideal non-invasive probes cannot be easily realized in
experimental situations. For this reason, the aim of the present
study is to go a step beyond the assumption of non-invasive probes
by considering probes that, while still weakly coupled to the
sample, introduce decoherence through inelastic scattering
processes, as well as inter-probe interference effects. Among other
interesting questions, our goal is to answer if features in the
behavior of the four-terminal resistance determined by non-invasive
probes, like Friedel oscillations, or a negative value of this
quantity,  are still possible when the coupling of the probes
becomes invasive.  We address these issues in the framework of
non-equilibrium Green functions and a perturbative treatment in the
coupling to the probes.

The work is organized as follows. In section II, we present the
model and the theoretical treatment to evaluate $R_{4t}$. In Section
III and IV, we present results for the clean wire, and the wire with
an impurity, respectively. Finally, we present a summary and
conclusions in Section V. Some technical details are presented in an
Appendix.

\section{Theoretical treatment}
\subsection{Model}

\begin{figure}
\includegraphics[width=0.9\columnwidth,clip]{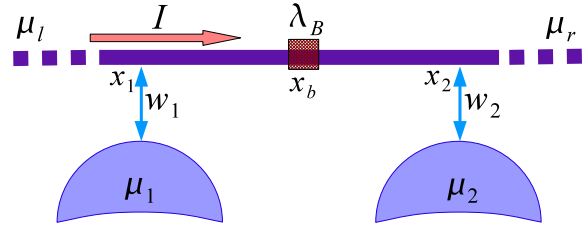}
\caption{\label{fig1}
 (Color online) Sketch of the setup: A voltage $V$ is imposed on a Luttinger
liquid, through the chemical potentials for the left and right
movers: $\mu_{r,l}=\mu \pm V/2$. Two voltage probes are connected at
the positions $x_1,x_2$. The corresponding chemical potentials
$\mu_{1,2}$ are fixed by the condition of zero current through the
contacts. An impurity is located at $x_b$.}
\end{figure}

We consider the setup of Fig.\ref{fig1}. As in Ref.
\onlinecite{ANS}, we use the following action to describe the full
system:
\begin{equation}
  S = S_{wire} + S_{imp}  + S_{res} +  S_{cont}
  \label{eq:action}
\end{equation}
where $S_{wire}$  corresponds to an infinite-length Luttinger wire and reads
\begin{widetext}
\begin{equation}
    S_{wire}  =  \int \, dx \, dt \, \Big \{
\psi^\dagger_l \Big[i(\partial_t - \partial_x) - \mu_l \Big] \psi_l
+\psi^\dagger_r
    \Big[i(\partial_t + \partial_x) - \mu_r \Big] \psi_r
    - g\left[\psi^\dagger_r \psi_r + \psi^\dagger_l \psi_l \right]^2
    \Big \},
    \label{eq:actionwire}
\end{equation}
\end{widetext}
with the first two terms corresponding to free spinless left and
right movers respectively and  $g$ is the e-e interaction in the
forward channel. We use units where $\hbar=k_B=1$. We also take the
Fermi velocity of the electrons $v_F=1$ and the electronic charge
$e=1$. The two chemical potentials $\mu_l = \mu - V/2$ and $\mu_r =
\mu + V/2$, for the left and right species, respectively, represent
a voltage bias $ V$  between the left and right ends of the wire,
which generates a current $I$.

The effect of the impurity is described by a backward scattering
term with strength $\lambda_B$ at a given position $x_b$:
\begin{equation}
  S_{imp} = \lambda_B \int \, dx \, dt \, \delta(x-x_b) \left[
    e^{-2ik_F x} \psi_r^\dagger \psi_l + \text{h.c.} \right].
  \label{eq:actionimp}
\end{equation}
We describe the voltage probes by
$S_{res}$, corresponding to non-interacting electrons with two
chiralities
\begin{eqnarray}
    S_{res}  & = &  \sum_{j=1,2} \int \, dy_j \, dt \, \Big \{
\chi^\dagger_{l_j} \Big[i(\partial_t - \partial_{y_j}) - \mu_{j} \Big] \chi_{l_j} \nonumber \\
& & +\chi^\dagger_{r_j}
    \Big[i(\partial_t + \partial_{y_j}) - \mu_{j} \Big] \chi_{r_j}
    \Big \},
    \label{eq:actionres}
\end{eqnarray}

The term $S_{cont}$ represents the tunneling between the reservoirs
and the wire,
\begin{eqnarray}
  S_{cont}  & = & \sum_{\substack{j=1,2 \\ \alpha=r,l, \beta_j = r_j, l_j}} \int
  \, dx \, dy_j \, w_j \delta(x-x_j)  \delta(y_j-y_j^0) \nonumber \\
& &
 \times  \left[e^{\mp i (k_F x + k_F^{(j)} y_j )} \psi^\dagger_\alpha
    \chi_{\beta_j} + \text{h.c.} \right]
  \label{eq:actioncont}
\end{eqnarray}
 The upper
and lower sign corresponds to $l$ and $r$, respectively, while $k_F$
and $k_F^{(j)}$ are the Fermi vectors of the wire and reservoirs,
respectively. Note that here we are assuming that the voltage probes
couple symmetrically to the left and right movers in the LL. This is
a natural assumption in the absence of magnetic fields and
spin-orbit interactions (Recall that we are considering a spinless
LL). Since our main interest is to discuss the effects originated in
the strength of the couplings, later, in Sections III and IV, as a
simplifying hypothesis, we will consider the same coupling for both
probes ($w_1=w_2$), although it is known that an asymmetrical
coupling ($w_1\neq w_2$) is sufficient to produce a negative
resistance\cite{takagaki-etal,ANS}.

The tunneling current from the probes to the wire is
\begin{equation}
  \label{eq:currdef}
  I_j = 2 \sum_{\alpha, \beta_j} \mbox{Re}
  \left[w_j G_{\alpha \beta_ j}^< (x_j, y_j^0; t, t ) \right],
\end{equation}
where
\begin{equation}
  G_{\alpha \beta_j}^< (x, y_j; t, t^{\prime}) =
  i <\chi_{\beta_j}^\dagger (y_j, t^{\prime}) \psi_\alpha(x, t) >,
  \label{eq:glesser1}
\end{equation}
is the lesser Green function involving degrees of freedom of the wire and
reservoirs.

\subsection{Green functions}
In addition to the lesser Green function defined in Eq.
(\ref{eq:glesser1}), we define the following retarded Green
functions:
\begin{eqnarray}
 & & G_{\alpha \beta}^R (x, x^{\prime}; t, t^{\prime}) = -i \Theta (t-t^{\prime})
   \langle \left \{ \psi_\alpha(x,t), \psi^\dagger_\beta(x^{\prime}, t^{\prime}) \right\} \rangle
  \label{eq:greenluttret1} \nonumber \\
& & G_{\alpha \beta j}^R (x, x^{\prime}; t, t^{\prime}) = -i \Theta
(t-t^{\prime})
   \langle \left \{ \psi_\alpha(x,t), \chi^\dagger_{\beta j} (x^{\prime}, t^{\prime}) \right\} \rangle,
  \label{eq:greenluttret2}
\end{eqnarray}
where the first one corresponds to degrees of freedom of the wire,
while the second one corresponds to degrees of freedom of the wire
and the $j$-th reservoir.

The evaluation of these Green functions implies the solution of the
Dyson equations. For the sake of simplicity in the notation, it is
convenient to  carry out the following gauge transformations
$\psi^{\dagger}_{l,r}(x) \rightarrow \psi^{\dagger}_{l,r}(x)  e^{\pm
i k_F x}$, $\chi^{\dagger}_{l_j,r_j}(y_j) \rightarrow e^{\pm i
  k_F^{(j)}  y_j}\chi^{\dagger}_{l_j,r_j} (y_j)$. The Dyson equation
for the retarded function reads
\begin{widetext}
\begin{eqnarray}
 & &\{  -i \partial_{t^{\prime}}  \pm k_F - \mu_{\beta} \pm i \partial_{x^{\prime}} \}     G_{\alpha
      \beta}^R (x, x^{\prime}; t, t^{\prime}) - \lambda_B G_{\alpha
     \overline{ \beta} }^R (x, x^{\prime}; t, t^{\prime}) \delta(x^{\prime}-x_b)
= \delta (t - t^{\prime}) \delta_{\alpha \beta} +  \nonumber \\
& &
\sum_j w_j G_{\alpha \beta_j}^R (x, y_j^0; t, t^{\prime} ) \delta (x^{\prime} - x_j )+ \sum_\gamma \int dx^{\prime \prime} dt^{\prime \prime} G_{\alpha
      \gamma}^R (x, x^{\prime \prime} ; t, t^{\prime \prime} )
    \Sigma_{\gamma \beta}^{int}(x^{\prime \prime} , x^{\prime} ; t^{\prime \prime}, t^{\prime})   \label{eq:dyson1a} \\
&  &\{  -i \partial_{t^{\prime} }  \pm k_F^{(j)} \pm i
\partial_{y_j} \}   G_{\alpha \beta_j}^R (x, y_j; t, t^{\prime} )  =
   w_j \sum_{\beta } G_{\alpha \beta}^R(x, x_j; t, t^{\prime} )
\delta(y_j-y_j^0),
  \label{eq:dyson1b}
\end{eqnarray}
\end{widetext}
where the upper and lower signs correspond to $\beta= l,r$ and
$\beta_j =l_j, r_j$, respectively, and $\overline{l}=r$,
$\overline{r}=l$, while $\Sigma_{\gamma \beta}^{int}(x^{\prime
\prime} , x^{\prime} ; t^{\prime \prime}, t^{\prime}) $ is the exact
self-energy due to the interaction term with coupling constant $g$.

Let us now notice that the operator
\begin{equation}
  -i \partial_{t^{\prime}} \pm k_F   \pm  i  \partial_{y_j^{\prime}} = \left\{ g_{\beta_j}^R
    (y_j, y_j^{\prime}; t, t^{\prime})\right\}^{-1},
  \label{eq:dysonreservoir}
\end{equation}
is the inverse of the retarded Green function corresponding to the
degrees of freedom $\beta_j$ of the reservoir $j$. Thus, Eq.
(\ref{eq:dyson1b}) can be expressed as follows
\begin{eqnarray}
    G_{\alpha \beta_j}^R (x, y_j^0; t, t^{\prime}) & = &
    w_j \sum_{\beta}  \int dt^{\prime \prime }
G_{\alpha \beta}^R (x, x_j; t, t^{\prime \prime } )  \nonumber \\
& &
   \times g_{\beta_j}^R (y_j^0, y_j^0; t^{\prime \prime } , t^{\prime } ) .
  \label{eq:gretarded}
\end{eqnarray}
Substituting the latter equation into Eq. (\ref{eq:dyson1b}) and
defining
\begin{eqnarray}
\Sigma_{\gamma \beta}^{res}(x, x^{\prime} ; t, t^{\prime})
& = & \sum_{j=1,2, \beta_j} \delta (x - x_j) \delta (x^{\prime} -x_j)
|w_j|^2  \nonumber \\
& & \times g_{\beta_j}^R(y_j^0, y_j^0; t, t^{\prime } ) ,
\end{eqnarray}
leads to
\begin{widetext}
\begin{eqnarray}
 & & \{  -i \partial_{t^{\prime}} \pm k_F -\mu_{\beta} \pm i \partial_{x^{\prime}} \} G_{\alpha
      \beta}^R (x, x^{\prime}; t, t^{\prime}) - \lambda_B G_{\alpha
     \overline{ \beta} }^R (x, x^{\prime}; t, t^{\prime}) \delta(x^{\prime}-x_b)
= \delta (t -
    t') \delta_{\alpha \beta} \nonumber \\
    & &
+ \sum_\gamma \int dx^{\prime \prime} dt^{\prime \prime} G_{\alpha
      \gamma}^R (x, x^{\prime \prime} ; t, t^{\prime \prime} )   \{
    \Sigma_{\gamma \beta}^{int}(x^{\prime \prime} , x^{\prime} ; t^{\prime \prime}, t^{\prime}) + \Sigma_{\gamma \beta}^{res}(x^{\prime \prime} , x^{\prime} ; t^{\prime \prime}, t^{\prime}) \}.
  \label{eq:dyson2}
\end{eqnarray}
\end{widetext}

The lesser Green function entering the expression for the
currents $I_j$ can be obtained  by means of Langreth rules from (\ref{eq:dyson1b}) \cite{lang},
according to which given $C^R(t,t^{\prime})= \int dt^{\prime \prime} A^R(t,t^{\prime \prime})
B^R(t^{\prime \prime}, t^{\prime})$ then $C^<(t,t^{\prime})= \int dt^{\prime \prime} [A^R(t,t^{\prime \prime})
B^<(t^{\prime \prime}, t^{\prime}) + A^<(t,t^{\prime \prime})
B^A(t^{\prime \prime}, t^{\prime})]$, being the advanced function $[B^A]^{\dagger}= B^R$ . Thus
\begin{widetext}
\begin{equation}
  G_{\alpha \beta_j}^< (x_j^0, y_j^0; t, t^{\prime}) = w_j \int dt^{\prime \prime}
  \left[ G_{\alpha \alpha}^<(x_j^0, x_j^0; t, t^{\prime \prime}) g_{\beta_j}^A
    (y_j^0, y_j^0; t^{\prime \prime}, t^{\prime}) +
G_{\alpha \alpha}^R(x_j^0, x_j^0; t, t^{\prime \prime})
    g_{\beta_j}^R (y_j^0, y_j^0; t^{\prime \prime}, t^{\prime}) \right]
  \label{eq:exactglesser}
\end{equation}
\end{widetext}
where $g^A=[g^R]^{\dagger}$ is the advanced Green function of the uncoupled reservoir.

So far all the equations are exact. The crucial step to obtain the
exact Green function by solving Dyson equations is the evaluation of
$\Sigma^{int}$, which corresponds to the fully dressed skeleton
diagram for the self-energy corresponding to the electron-electron
interaction, also taking into account the coupling to the two
additional reservoirs as well as the backward impurity. We now
introduce the following approximation for the limit of weak coupling
to the reservoirs and the impurity such that $w_j \ll g$ and
$\lambda_B \ll g$:
\begin{equation}
 \Sigma_{\gamma \beta}^{int}(x , x^{\prime} ; t, t^{\prime})
\simeq \Sigma_{\gamma \beta}^{Lutt}(x , x^{\prime} ; t, t^{\prime}),
\label{selfaprox}
\end{equation}
where
\begin{eqnarray}
& & \Sigma_{\gamma \beta}^{Lutt} (x , x^{\prime} ; t, t^{\prime})
 =  [ G^{R,Lutt} (x , x^{\prime} ; t, t^{\prime}) ]^{-1}_{\gamma \beta}   -
\nonumber  \\
& & \{ -i \partial_{t^{\prime} } \pm  k_F \pm i \partial_{x^{\prime} } \} \delta_{\gamma \beta} ,
\end{eqnarray}
is the self-energy of the infinite Luttinger wire without impurity
and uncoupled from the reservoirs, while $G^{R, Lutt}(x , x^{\prime}
; t, t^{\prime})$ is the ensuing retarded Green function. The
approximation (\ref{selfaprox}) implies the evaluation of the
self-energy associated to e-e interaction by neglecting vertex
corrections due to the escape to the reservoirs and due to the
scattering with the impurity.  This approximation is adequate only
in the limit of small $w_j$ and $\lambda_B$.

Under this approximation in the e-e self-energy and performing a Fourier transform
with respect to $t-t^{\prime}$, Eq. (\ref{eq:dyson2}) can be expressed as
follows
\begin{widetext}
\begin{equation}
  G_{\alpha \beta}^R (x, x^{\prime}; \omega) = G_{\alpha \beta}^{R, Lutt}
  (x, x^{\prime}; \omega) + \sum_j G_{\alpha \alpha}^R (x, x_j, \omega) \Sigma^{res}_{\alpha \beta}(x_j,x_j;\omega)
  G_{\beta \beta}^{R, Lutt} (x_j, x^{\prime}; \omega)+ G_{\alpha \alpha}^R (x, x_b, \omega)
\lambda_B  G_{\beta \beta}^{R, Lutt} (x_b, x^{\prime}; \omega)
  \label{eq:approx3}
\end{equation}
\end{widetext}
This  equation allows  for the evaluation of the retarded Green
function. In what follows, we solve it at the lowest order in the
backscattering term $\lambda_B$ and up to ${\cal O}( w_j^2)$, in the
coupling to the voltage probes. We recall that ideal non-invasive
probes correspond to keeping only up to ${\cal O}(w_j)$. It is important to notice
that in the limit of vanishing Coulomb interaction ($g=0$), the above equation leads
the exact retarded Green function of the problem.

\subsection{Currents}

Substituting Eq. (\ref{eq:exactglesser}) in the definition of the
current \eqref{eq:currdef}, we get the following exact equation for
the current flowing through the contact between the $j$-th reservoir
and the wire
\begin{widetext}
\begin{equation}
    I_j = -2 \sum_{\alpha \beta, j} \mbox{Re} \left\{ w_j^2 \int \frac{d\omega}{2\pi}
      \left[ G_{\alpha \alpha}^< (x_j^0, x_j^0; \omega)
        g_{\beta_j}^A(r_j^0, r_j^0; \omega) +
        G_{\alpha \alpha}^R (x_j^0, x_j^0; \omega)
        g_{\beta_j}^< (r_j^0, r_j^0; \omega)\right]\right\}.
  \label{eq:exactcurrent}
\end{equation}
\end{widetext}

Making use of the assumption of weak coupling between the probes and
the wire and weak amplitude in the back scattering term induced by
the impurity, we evaluate the Green functions $G_{\alpha \alpha}^<
(x_j^0, x_j^0; \omega) $ and $ G_{\alpha \alpha}^R (x_j^0, x_j^0;
\omega) $ perturbatively up to ${\cal O}(w_j^2)$ and ${\cal
O}(\lambda_B)$. Concretely, this implies solving (\ref{eq:approx3})
as
\begin{widetext}
\begin{eqnarray} \label{retaprox}
  G_{\alpha \beta}^R (x, x^{\prime}; \omega)  &\simeq & G_{\alpha \beta}^{R, Lutt}
  (x, x^{\prime}; \omega) + \sum_j G_{\alpha \alpha}^{R, Lutt}  (x, x_j, \omega) \Sigma^{res}_{\alpha \beta}(x_j,x_j;\omega)
  G_{\beta \beta}^{R, Lutt} (x_j, x^{\prime}; \omega)\nonumber \\
& & + G_{\alpha \alpha}^{R, Lutt} (x, x_b, \omega) \lambda_B
G_{\beta \beta}^{R, Lutt} (x_b, x^{\prime}; \omega) ,
\end{eqnarray}
\end{widetext}
while the lesser counterpart can be derived from (\ref{retaprox}) by
recourse to Langreth rules (see above Eq. (\ref{eq:exactglesser})).
The explicit expression for $ G_{\alpha \alpha}^{R, Lutt} (x, x_j,
\omega)$ is given in Appendix A. After some algebra, the currents
through the contacts can be expressed as follows
\begin{equation}
I_j= I^{(1)}_j+ I^{(2)}_j,
\end{equation}
where $I^{(1)}_j \propto |w_j|^2$, $I^{(2)}_j \propto |w_j|^4$,
being

\begin{widetext}
\begin{eqnarray}
  I_j ^{(1)} &=& 2 w_j^2 \sum_{\alpha = l,r} \int_{-\infty}^\infty
  \frac{d\omega}{2\pi}  \Bigg \{  G^{<, Lutt}_{\alpha \alpha} (x_j, x_j, \omega)
    g_j^>(\omega) - G^{>,Lutt}_{\alpha \alpha} (x_j, x_j, \omega)
    g_j^<(\omega)  \nonumber \\
   & & + \lambda_B
     \Big[ [G_{\alpha \alpha}^{<,Lutt} (x_j - x_b; \omega) \left( G_{\overline{\alpha}
          \overline{\alpha}}^{R,Lutt} (x_j - x_b; \omega) \right)^\ast +
          G_{\alpha \alpha}^{R,Lutt}(x_j - x_b; \omega ) G_{\overline{\alpha} \overline{\alpha}}^{<,Lutt}(x_b - x_j) ]
    g_j^>(\omega) - \nonumber \\
     \qquad &-& [G_{\alpha \alpha}^{>,Lutt} (x_j - x_b; \omega) \left(
        G_{\overline{\alpha} \overline{\alpha}}^{R,Lutt} (x_j - x_b; \omega) \right)^\ast + G_{\alpha
        \alpha}^{R,Lutt}(x_j - x_b; \omega ) G_{\overline{\alpha} \overline{\alpha}}^{>,Lutt}(x_b - x_j) ]
    g_j^< (\omega) \Big] \Bigg \} \\
  I_j^{(2)} &=& 4 w_j^2 \sum_{\substack{i=1,2\\\alpha, \beta = r, l}} w_i^2
  \int_\infty^\infty \frac{d\omega}{2 \pi}  \Big\{G_{\alpha \alpha}^{R, Lutt}(x_j, x_i, \omega)
  G_{\beta \beta}^{A,Lutt} (x_i, x_j, \omega) \left[
    g_j^>(\omega) g_i^<(\omega) - g_j^<(\omega) g_i^>(\omega) \right] + \nonumber
  \\\label{eq:wickaproximado}
  &+& G_{\alpha \alpha}^{R, Lutt}(x_j, x_i, \omega) g_i^R(\omega)
  \left[G^{<,Lutt}_{\beta \beta } (x_i, x_j, \omega) g_j^>(\omega) -
    G^{>,Lutt}_{ \beta \beta} (x_i, x_j, \omega) g_j^<(\omega) \right] + \\
  &+& G_{\alpha \alpha}^{A,Lutt}(x_i, x_j, \omega) g_i^A(\omega)
  \left[G^{<,Lutt}_{ \beta \beta} (x_j, x_i, \omega) g_j^>(\omega) -
    G^{>,Lutt}_{\beta \beta} (x_j, x_i, \omega) g_j^<(\omega) \right] \Big\}
  \nonumber
  \label{eq:currentimpu}
\end{eqnarray}
\end{widetext}
The term $I^{(1)}_j$ corresponds to the limit of ideal non-invasive
contacts considered in Ref. \onlinecite{ANS}. It is derived by
dropping the second term in Eq. (\ref{retaprox}) and  the ensuing
term in the lesser counterpart. This leads to the exact solution
${\cal O}(w_j)$ of (\ref{eq:dyson2}) and (\ref{eq:exactglesser}) for
$\lambda_B=0$. In the second-order solutions (\ref{retaprox}) we
have introduced the additional approximation of neglecting vertex
corrections $\propto w_j^2$ and $\propto \lambda_B$ in the
evaluation of the many-body self-energy $\Sigma^{int}$. Notice that
the two probes are completely uncorrelated within the
''non-invasive'' component $I^{(1)}$. In the higher order
contribution $I^{(2)}$ it is possible to distinguish two kinds of
terms. On one hand, those $\propto w_j^4$, account for the effect of dephasing
and resistive behavior induced by the inelastic scattering processes
due to the coupling to the reservoirs. On the other hand, terms
$\propto w_1^2 w_2^2$ describe interference effects between the two
probes.

It is now convenient to express the lesser and greater Green
function in terms of spectral functions:
\begin{eqnarray}
  G_{\alpha, \alpha}^{<,>,Lutt} (x_1 - x_2; \omega+ \mu_{\alpha}) &=& \lambda^{<,>}_{\alpha}(\omega)
  \rho_\alpha(x_1 - x_2; \omega) , \nonumber \\
  g_j^{<, >} (\omega) &=&\lambda^{<,>}_{j}(\omega) \rho_j(\omega)
\end{eqnarray}
with $\lambda^<_{\alpha, j}(\omega) = i n_{F,\alpha,j} (\omega )$,
$\lambda^>_{\alpha,j} (\omega)= -i [1 -n_{F,\alpha,j}(\omega) ]$,
being $n_{F, \alpha } (\omega) = 1/(e^{(\omega \pm  V/2)/T} +1 )$,
the Fermi function where the upper and lower signs corresponds,
respectively to the right and left movers of the wire, and $n_{F,j }
(\omega) = 1/(e^{(\omega - \mu_j)/T} +1 )$, $\mu_j$ being the
chemical potentials of the electrons in the $j$ reservoir, relative
to the mean chemical potential $\mu$ of the wire. $T$ is the
temperature, which we assume to be the same for the wire and the
probes, while $\rho_{\alpha}(\omega) =i G^{R, Lutt}_{\alpha
\alpha}(x_1- x_2;\omega) -
 i[G^{R, Lutt}_{\alpha \alpha}(x_2- x_1;\omega)]^*$ is the spectral function for the $\alpha$ movers in the
Luttinger model,
 and $\rho_j(\omega) = -2 \mbox{Im}[g^R_j(\omega)]$, is the spectral density of the $j$ probe.
 Replacing in
\eqref{eq:wickaproximado} and \eqref{eq:currentimpu}, the full
expression for the current reads:
\begin{widetext}
\begin{eqnarray} \label{curj}
    I_j &= & 2 w_j^2 \sum_{\alpha, \beta = l,r} \int_\infty^\infty
    \frac{d\omega}{2 \pi} [n_{F, \alpha} (\omega ) -
      n_{F,j} (\omega) ] \rho_j (\omega) \{ \rho_{\alpha} (0, \omega)
\left[1 + 2 w_j^2 \mbox{Re}
      \left(G_{\beta}^{R,Lutt} (0, \omega) g_j^R(\omega) \right) \right] +  \nonumber \\
    & & + 2 \lambda_B \mbox{Re} \left[ \rho_{\alpha} (x_j - x_b; \omega)
      \left( G_{\bar \alpha}^{R, Lutt} (x_j - x_b; \omega) \right)^\ast \right] +  \qquad  2 w_{\bar j}^2 \mbox{Re}
        \left[ \rho_\alpha (x_j - x_{\bar j}; \omega)^\ast
      G_{\beta}^{R, Lutt}  (x_j - x_{\bar{j}}; \omega) g_{\bar{j}}^R (\omega) \right]  \}  \nonumber \\
    & & + 2 w_j^2 w_{\bar j}^2 \sum_{\alpha, \beta = l,r}
    \int_\infty^\infty \frac{d\omega}{2 \pi} \left[n_F(\omega -
      \mu_{\bar j}) -
      n_F(\omega - \mu_j)\right] \rho_j(\omega) \rho_{\bar j}(\omega)
 G_{\beta}^{R, Lutt} (x_j - x_{\bar j}; \omega) \left (G_{\alpha}^{R, Lutt}
 (x_j - x_{\bar j}; \omega)\right)^\ast,
  \label{eq:fullcurrent}
\end{eqnarray}
\end{widetext}
where we use the notation $j, \bar j$ such that $\bar 1=2$ and $\bar
2=1$.

\subsection{ Voltage drop and four-terminal resistance}
The chemical potentials $\mu_j$ in the expressions of the previous
subsection must be set to satisfy the condition of local
electrochemical equilibrium between the probes and the wire. This
implies vanishing flows $I_j=0$, $j=1,2$, with the currents defined
in Eq. (\ref{curj}), and the two chemical potentials $\mu_j$ must
satisfy these constraints. In the case of non-invasive probes, the
two probes are completely uncorrelated, and the problem can be
reduced to that of the wire coupled to a single probe, which senses
the local chemical potential of the wire. Instead, in the present
case, we have to solve a system of two non-linear equations to
calculate $\mu_1$ and $\mu_2$, from where the voltage drop $\Delta
V= \mu_1 - \mu_2$ between the  points $x_1$ and $x_2$ of the wire
coupled to the two probes can be evaluated. This voltage drop
contains not only information of the scattering processes in the
wire that are independent of the coupling to the probes, but also of
inelastic scattering processes and interference effects introduced
by the probes themselves. The four-terminal resistance can be
evaluated from Eq. (\ref{r4t}) and the ratio between the
four-terminal and two terminal resistance $R_{2t}=V/I$ results
\begin{equation} \label{ratio}
\frac{R_{4t}}{R_{2t}}= \frac{\Delta V}{V}.
\end{equation}
The two chemical potentials are evaluated numerically from Eq. (\ref{curj}), with the Green functions
given in Appendix A.

\section{Results without impurity}
In this section we show results for the ratio between
$R_{4t}/R_{2t}$ in the case of $\lambda_B=0$. It is important to
mention that in the limit of non-invasive probes, this ratio
vanishes identically under this case, and all the  features in the
behavior of the resistance discussed in this section are solely due
to the invasive nature of the probes.

We characterize the strength of e-e interactions with the parameter
$K=(1+\frac{2g}{\pi})^{-1/2}$. The limit of non-interacting
electrons corresponds to $K=1$ while typical values of $K$
(experimentally determined in transport measures in nanodevices
\cite{K-values}) are in the range $0.25 < K < 0.75$.

\begin{figure}
    \includegraphics[width=0.9\columnwidth,clip]{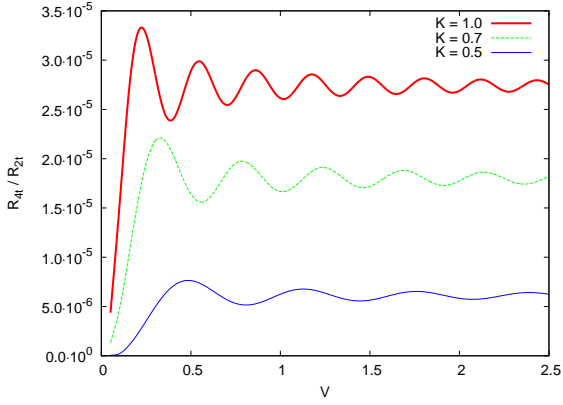}
    \caption{(Color online)\label{fig2}$R_{4t}/R_{2t}$ as a function of the
    voltage $V$, for different values of the e-e interaction strength $K$. The positions of the
    probes are $x_1 = -10$ and $x_2 = 10$, and the strength of the
    couplings are $w_1 = w_2 = 0.1$.  The temperature is $T=0$.}
\end{figure}

\begin{figure}
    \includegraphics[width=0.9\columnwidth,clip]{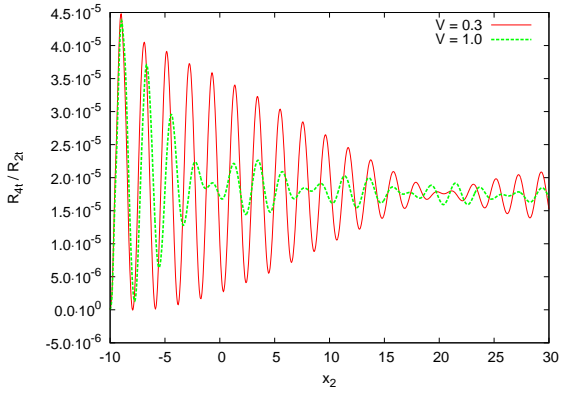}
    \caption{(Color online)\label{fig3} $R_{4t}/R_{2t}$ as a function of the position of the second probe
    $x_2$, given the first probe fixed at $x_1 = -10.$. The strength of the e-e interaction
    is $K=0.7$, and the couplings $w_1 = w_2 = 0.1$.}
\end{figure}

\begin{figure}
    \includegraphics[width=0.9\columnwidth,clip]{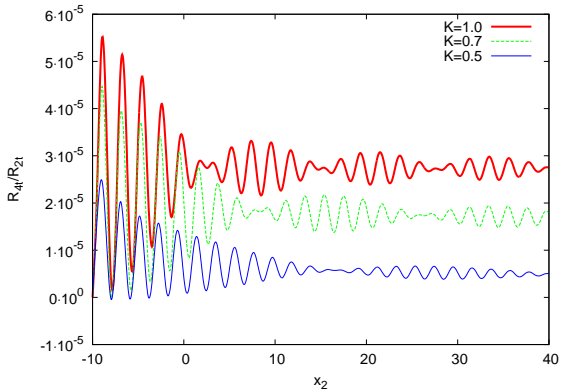}
    \caption{(Color online)\label{fig4} $R_{4t}/R_{2t}$ as a function of the position of the second probe
    $x_2$, given the first probe fixed at $x_1 = -10.$, for different values of the e-e interaction strength $K$.
    We set $T=0$, $w_1 = w_2 = 0.1$ and $V=1$.}
\end{figure}

\begin{figure}
    \includegraphics[width=0.9\columnwidth,clip]{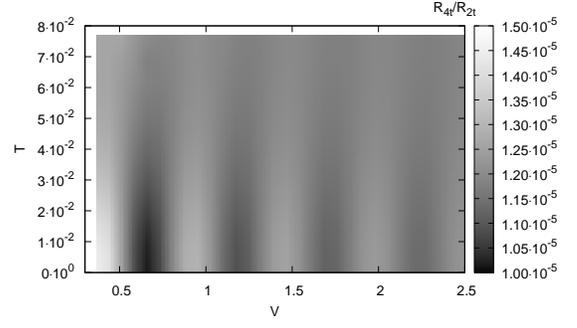}
    \caption{(Color online)\label{fig5}$R_{4t}/R_{2t}$ as a function of the
    voltage $V$ and the temperature $T$, for $K=0.7$. The positions of the
    probes are $x_1 = -10.$ end $x_2 = 10.$, and the strength of the
    couplings are $w_1 = w_2 = 0.1$. }
\end{figure}
\begin{figure}
    \includegraphics[width=0.9\columnwidth,clip]{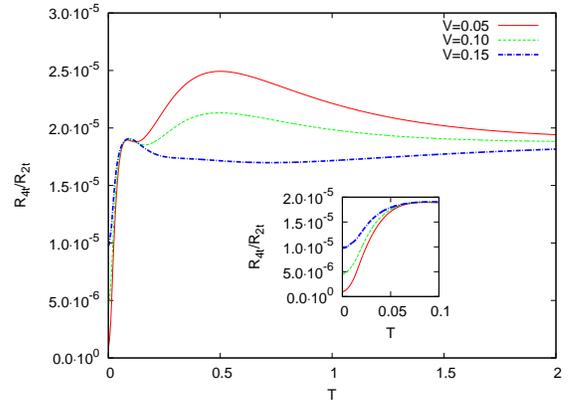}
    \caption{(Color online)\label{fig6}$R_{4t}/R_{2t}$ as a function of the temperature
    $T$, for $K=0.7$ and three values of $V$. The positions of the
    probes are $x_1 = -10.$ end $x_2 = 10.$, and the strength of the
    couplings are $w_1 = w_2 = 0.1$. Inset: low temperature regime.}
\end{figure}

Results for   $R_{4t}/R_{2t}$ as a function of the bias voltage $V$,
for  the probes connected at two fixed positions and different
values of the e-e interaction $K$  are shown in Fig. \ref{fig2} In
order to gain insight on the behavior of the ratio between
resistances, let us notice that for vanishing bias voltage $V$, the
voltage drop $\Delta V$ and thus $R_{4t}/R_{2t}$ should be also
vanishing. It is, therefore, not surprising that for low enough $V$,
$R_{4t}/R_{2t}$ displays a power law behavior as a function of $V$,
\begin{equation}
\frac{R_{4t}}{R_{2t}} \propto V^{2 \gamma + 1}.
\end{equation}
The exponent $\gamma$ is related to the Luttinger parameter as
$\gamma= (K+K^{-1}-2)/4$. The latter result can be rather
straightforwardly derived from an expansion of $I_j$ for low $V$. On
the other hand, a classical ohmic-like resistive behavior implies a
constant value of $R_{4t}/R_{2t}$. In Fig. \ref{fig2}, it can be
seen that such a behavior is approximately attained when the bias
voltage overcomes a value $V_c$, which satisfies
 $e V_c \approx \hbar v/( 2 K \mid x_1-x_2 \mid )$. \cite{note}  This energy scale can be understood by noticing that
$v/K$ is the plasmon velocity along the wire and $\tau_p = 2 K  \mid
 x_1-x_2 \mid / v $ is the time that these excitations take for a round
trip between the probes. The latter defines the characteristic time
for the  inelastic back-scattering processes. Notice, that although
the Luttinger wire is an elastic system, where electrons propagate
ballistically, the coupled voltage probes  act as a dissipative
bath. In fact, it is precisely the coupling to reservoirs the
mechanism usually followed in the literature (see, for example Refs.
\onlinecite{dampast} and \onlinecite{caz}) in order to model Ohmic
dissipation. In the present case, assuming that the bias is applied
from left to right, the Fermi energy of right-moving electrons is an
amount $eV$ higher
 than that of the left-moving ones.
Then, the energy associated to the crossover voltage $e V_c$ corresponds
to the energy dissipated in the contacts, in a process in which an
electron with the Fermi energy $\mu_R$
 travels with velocity $v/K$ from the left probe, connected
at $x_1$, to the right one, connected at $x_2$, it is backscattered
at $x_2$ and comes back to $x_1$ with a Fermi energy $\mu_L$.
An estimate for the energy transfer involved in the
dissipative process is, precisely, $\hbar \tau_p^{-1}$. The above
argument can be easily reconstructed for the case of a bias with
opposite sign, in which case, the energy is inelastically
transferred from right to left movers. Notice that in any case, the
energy dissipated at the contacts is associated to a voltage drop
that has the same direction as the external bias, as is expected for
a classical Ohmic-like process.
 Interestingly, $2 \tau_p^{-1}$ is
 equivalent to the ballistic frequency defined in Ref. \onlinecite{dolc}
for an interacting Luttinger wire of finite length connected to reservoirs.

 To summarize, for a given separation  $\mid
 x_1-x_2 \mid $ between the two probes, $V_c$ defines the crossover voltage
 for which inelastic back-scattering processes between the two points become active.
 Notice that the low voltage regime $V<V_c$ so defined,
depends on the e-e interaction strength $g$, being wider for
stronger $g$ (lower $K$). In general, the effect of this interaction
is to decrease the resistance.
 A closer analysis of  Fig. \ref{fig2} for $V>V_c$ reveals that $R_{4t}/R_{2t}$ as a function of the bias voltage $V$
 displays oscillatory features. This can be naturally interpreted
 as the consequence of interference effects between the two probes.
From fits of the numerical data, we found that they can be very well
reproduced by a function of the form:
\begin{equation}
\frac{R_{4t}}{R_{2t}} \approx A + B \sin(KV(x_2-x_1))/V^{2\gamma
+1}, \;\;\;\; V>V_c,
\end{equation}
with $A$ and $B$ depending on $K$ while
proportional to $w^2$, although we have not derived this result analytically
from Eq. (\ref{curj}). It is anyway interesting that a similar resistive behavior
is obtained in a Luttinger wire of finite length in the presence of back-scattering
processes (see Refs. \onlinecite{dolc,pug}). Another interesting observation is that the
saturation value $A$ decreases for increasing electron-electron interactions, This
indicates that the latter tend to screen the inelastic scattering processes introduced
by the coupling to the probes.

In Fig. \ref{fig3} we show the
behavior of the ratio between resistances with the position of one
of the probes kept fixed while the position of the second one is
moved along the wire. This pattern reveals that the functional
behavior is
\begin{equation}
\frac{R_{4t}}{R_{2t}} \approx A + B \sin (2 k_F (x_2-x_1))
\sin(KV(x_2-x_1))/V^{2\gamma+1},
\end{equation}
within the high $V>V_c$ voltage regime, corresponding, respectively,
to solid and dashed lines in the Figure. The $2 k_F$ modulation
resembles the behavior found in the voltage profile of non-invasive
probes in a system with an impurity, which is observed both in
non-interacting \cite{but89,pern,grambu} and interacting systems \cite{ANS}. In
those cases the origin is the occurrence of interference in the
electronic wave packet generated by the back-scattering processes
that take place at the impurity. In the present case, the
interference is originated by scattering processes at the probes.
Unlike the behavior for non-invasive probes, in our case the voltage
drop induced by the probes has the same sign as the applied external
voltage. This means that the four-terminal resistance for invasive
probes in a clean wire is always positive, in spite of the
Friedel-like $2 k_F$ oscillations. This is in strong  contrast to
the case of non-invasive probes, where these oscillations provide a
mechanism for $R_{4t} <0$. Fig. \ref{fig4} illustrates the same
situation but for fixed voltage and varying $K$. One sees that, in
general, larger values of the e-e interactions produce smaller
values of $R_{4t}/R_{2t}$. Then we conclude that, although one
cannot have negative values of the four terminal resistance in the
absence of impurities, e-e interactions tend to facilitate that
possibility.

In Fig. \ref{fig5} we show the effect of the temperature in the
behavior of  $R_{4t}/R_{2t}$. It is clear that, as the temperature
increases, the oscillations discussed in Fig. \ref{fig2} within the
high voltage regime, tend to be wiped out and the resistance evolves
to a constant value. This behavior is depicted in more detail in
Fig. \ref{fig6}, where we display $R_{4t}/R_{2t}$ as function of $T$
for three different values of the bias voltage $V$. In analogy with
the previously discussed behavior found at $T=0$, as function of $V$
(Fig.\ref{fig2}), there is a crossover temperature $T_c \approx \hbar
v/(2 K \mid
 x_1-x_2 \mid)$ which allows to distinguish low and high temperature regimes.
For low temperatures ($T<T_c$), we have verified that the ratio
between resistances behaves as
\begin{equation}
\frac{R_{4t}}{R_{2t}} \simeq a + b\, T^{2 \gamma + 1},
\end{equation}
where $a$ and $b$ depend on $V$. For high temperatures
$R_{4t}/R_{2t}$ tends to a constant value. As the temperature
increases, coherence tends to disappear. For this reason, no
signature of the oscillatory behavior observed in Fig. \ref{fig2} is
found here. The interplay between $T$ and $V$ gives rise to the
possible occurrence of a global maximum of the $R_{4t}/R_{2t}$. The
value of temperature that corresponds to this maximum, when it is
present, depends on $K$ and $\mid x_1-x_2 \mid$.

\section{Results with impurity}
In this section we analyze the behavior of $R_{4t}/R_{2t}$ at $T=0$,
for a wire with an impurity with backscattering strength
$\lambda_B$. In the case of non-invasive probes, the local voltage
displays $2k_F$ Friedel-like oscillations with constant amplitude
for non-interacting electrons \cite{but89,pern,grambu}, and with modulated
amplitude in the case of and interacting wire. \cite{ANS}
\begin{figure}
    \includegraphics[width=0.9\columnwidth,clip]{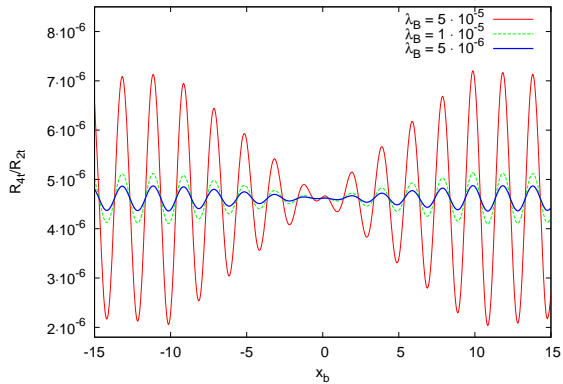}
    \caption{(Color online)\label{fig7} $R_{4t}/R_{2t}$ as a function of the position of the impurity
    $x_b$, for probes connected at $x_1 = -10.$ and $x_2=10$. The bias
voltage is $V=0.5$, the strength of the e-e interaction
    is $K=0.7$, and the couplings $w_1 = w_2 = 0.1$. Different plots correspond to different strengths
of the backscattering term representing the impurity.}
\end{figure}

Figure \ref{fig7} shows $R_{4t}/R_{2t}$ for the probes connected at
fixed positions, as a function of the position of the impurity
$x_b$. Friedel-like oscillations with period $2k_F$ are identified,
with an increasing amplitude for increasing back-scattering
strength. As in the case of non-invasive probes, the amplitude is
modulated for interacting electrons, the local voltage achieving the
highest amplitudes at the position of the impurity. Unlike the case
of non-invasive probes, the oscillations take place with respect to
a constant non-vanishing value, which is determined by the degree of
coupling of the probes.  For the parameters shown in the figure,
$R_{4t}$ is always a positive quantity.

\begin{figure}
    \includegraphics[width=0.9\columnwidth,clip]{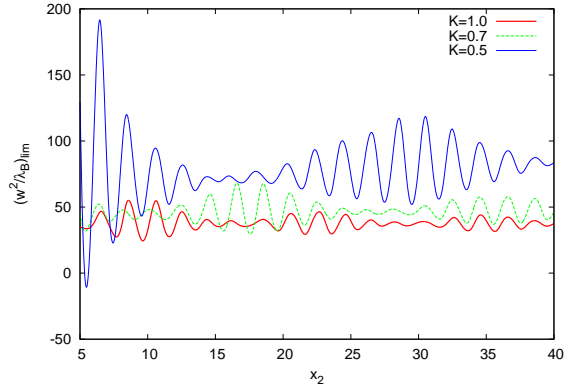}
    \caption{(Color online) \label{fig8}The ratio $(w_j^2/\lambda_B)_{lim}$ at which $R_{4t}=0$ for $x_1=x_b=-10$,
as a function of the position of the second probe $x_2$. Different
plots correspond to different e-e interaction parameters $K$. The
applied voltage is $V=0.5$.}
\end{figure}
Besides interference effects, it is clear that the coupling of the
probes generates classical resistive behavior through inelastic
scattering processes, while the elastic scattering induced by the
impurity induce Friedel oscillations. The first type of processes
takes place with a strength $\propto w_j^2/\Lambda^{\prime}$, where
$\Lambda^{\prime}$ is the bandwidth of the reservoirs, while the
second one takes place with a strength $\lambda_B$. The two
mechanisms are competitive regarding the possibility of having
$R_{4t} <0$. In Fig. \ref{fig8} we analyze, precisely, this
possibility. To this end, we have fixed the first probe at the
position $x_1=x_b=-10$, where the minimum $R_{4t}/R_{2t}$ is
achieved, considering different positions for the second probe
$x_2$. For each of these configurations we then vary the ratio $
w_j^2/\lambda_B$, to define $(w_j^2/\lambda_B)_{lim}$, at which
$R_{4t}/R_{2t}=0$. The corresponding results are plotted in the
figure for different e-e interactions. For $w_j^2/\lambda_B >
(w_j^2/\lambda_B)_{lim}$ the ratio $R_{4t}/R_{2t}$ is positive for
any value of $x_b$. On the other hand, the condition
$w_j^2/\lambda_B < (w_j^2/\lambda_B)_{lim}$ defines the values of
coupling strength for which a negative four terminal resistance is
possible, depending on the position of the impurity.

A very interesting and subtle issue that is also revealed by our
analysis concerns the role of e-e interactions in the possible
occurrence of a negative four-terminal resistance. Based on the
results obtained for non-invasive probes
\cite{but89,pern,grambu,ANS} one would expect that e-e interactions
oppose to such possibility, owing to the fact that for stronger
interactions (smaller values of $K$) the amplitude of the
oscillations coming from the presence of the impurity diminishes.
However, in the present case this effect competes with the global
"upward" shift coming from the contribution of $I^{(2)}_j$. In other
words, as already pointed out in Section III, the weak invasiveness
of the probes, which in our formulation is contained in $I^{(2)}_j$
produces a voltage drop that has the same sign of the bias $V$. It
turns out that the magnitude of such a shift also depends on $K$,
and it decreases for increasing interactions (decreasing $K$), as
shown in Figures (\ref{fig2}) and (\ref{fig6}). The combination of
these two effects gives rise to the result depicted in
Fig.(\ref{fig8}), where one sees that for sufficiently separated
probes, e-e interactions facilitate the occurrence of a negative
four-terminal resistance.

\section{Summary and Conclusions}
We have analyzed the behavior of the four terminal resistance in a
biased quantum wire with an impurity. We have modeled the wire by an
infinite-length Luttinger wire where the bias voltage is represented
by different chemical potentials for the left and right movers, and
the impurity by a backscattering term. We have also introduced
models for the probes, which consist in reservoirs of
non-interacting electrons. These systems are locally weakly coupled
to the wire and have chemical potentials satisfying the conditions
of vanishing electronic currents between the reservoirs and the
wire. The difference between the so determined chemical potentials
defines the voltage drop, from where the ratio between the
four-terminal and two-terminal resistance can be calculated. We have
solved the problem within perturbation theory in the impurity
strength and the tunneling parameter defining the coupling between
the probes and the wire, within the framework of non-equilibrium
Green functions formalism. We have neglected vertex corrections in
the self-energy for the e-e interaction associated to inelastic
scattering processes due to the escape to the leads and elastic
scattering processes at the impurity. Since we have assumed that
these two parameters are small enough, the latter is expected to be
a reliable approximation.

We have analyzed the voltage drop beyond the non-invasive assumption
for the coupling of the probes to the wire. That is, we have
studied, not only the voltage drop originated by elastic scattering
processes along the wire, but also the effects introduced by the
coupling to the probes, itself. We have shown that the inelastic
scattering processes due to the invasive coupling of the probes
induce a voltage drop with a power law behavior as a function of the
bias voltage for low values of this parameter, with an exponent
determined by the e-e interaction. In the limit of non-interacting
electrons, this reduces to a linear dependence as a function of the
bias voltage. This behavior has classical and quantum features,
since the voltage drop is always in the same sense of the applied
voltage but displays a pattern of oscillations indicating quantum
interference between the two probes. These features, are, however,
screened as the e-e interaction increases. In our calculations, we
have considered an infinite wire. However, the separation between
the probes sets a natural length scale in the problem, which
determines the crossover value of the bias voltage for which
inelastic scattering processes become active. In the case of an
interacting wire with finite length, we expect that our results
remain valid provided the length of the wire is much larger than the
separation between the probes. In the presence of an impurity, the
elastic backward scattering processes and $2k_F$ oscillations
detected by non-invasive probes \cite{ANS} are superimposed to the
inelastic processes introduced by the probes.

Our results have an important outcome in relation to experimental
measurements of four-terminal resistance in real systems. That is,
for invasive probes, elastic effects like those generated by
backscattering processes by impurities can still lead to a voltage
drop that opposes to the applied voltage, giving rise to a negative
four-terminal resistance. However, the amplitude for these processes
must be high enough in order to overcome the classical resistive
effect introduced by the probes.

As far as the e-e interaction effects are concerned, they play a
fundamental role in the calculated magnitudes. For higher e-e
interaction, the oscillations amplitude coming from the impurity
decreases. The amplitude of the global shift coming from the
interaction of the probes also decreases for stronger interactions.
We have shown that if the separation of the probes is large enough,
the possibility of measuring a negative resistance increases for
stronger interactions.

\section{Acknowledgments}
We acknowledge support from CONICET and  ANPCYT, Argentina, as well
as UBACYT and the J. S. Guggenheim Memorial Foundation (LA).

\appendix

\section{Green functions and spectral functions for the Luttinger
wire and the reservoirs}

We can follow the procedure of Ref. \onlinecite{gflut} to evaluate
the retarded Green functions of the Luttinger wire and calculate the
spectral density from $ \rho_{\alpha}(x, x^{\prime}; \omega) =
i[G^{R, Lutt}_{\alpha}(x-x^{\prime}  ; \omega) - G^{R,
Lutt}_{\alpha} (x^{\prime} -x  ; \omega)^*]$. The result is
\begin{eqnarray}
  \rho_{\alpha}(x, x^{\prime}; \omega + \mu_{\alpha})  & = &
  C \exp \left[\mp i \left(\frac{\omega}{v} - k_F \right) x \right]
  |\omega|^{2\gamma}
\nonumber \\
& &
\times \phi (\gamma, 2 \gamma + 1, \pm 2i x \omega/v),
  \label{eq:spectraldensity}
\end{eqnarray}
where $\phi$ is Kummer's Hypergeometric function, $\gamma=
(K+K^{-1}-2)/4$ and $v=1/K$. In order to perform numerical
calculations, we introduce an energy cutoff $\Lambda$ such that
 $\rho(x, \omega+\mu) \rightarrow \rho(x, \omega+\mu)
\Theta(\Lambda - |\omega|)$.  $C$ is a normalization constant, which
can be evaluated by the normalization condition
\begin{equation}
  \int \frac{d\omega}{2\pi} \rho(x, \omega)\big|_{x=0} = 1
  \label{eq:normalization}
\end{equation}
The retarded and advanced green function are then calculated using the
Kramers-Kronig relations
\begin{eqnarray}
    G^{R, Lutt} (x,\omega) &= & \int \frac{d \omega^{\prime}}{2\pi}
    \frac{\rho(x, \omega^{\prime})}{\omega - \omega^{\prime} + i \eta} =  \nonumber \\
    &= & \int \frac{d \omega^{\prime}}{2\pi} \frac{\mbox{Re}[ \rho(x, \omega^{\prime}) ] + i \mbox{Im}
      [\rho(x, \omega^{\prime}) ] }
    {\omega - \omega^{\prime} + i \eta} = \\
    &=  & P \int \frac{d \omega^{\prime}}{2\pi} \frac{\rho(x, \omega^{\prime})} {\omega -
      \omega^{\prime}} - \frac{i}{2} \rho(x, \omega).
  \label{eq:kramerskronig}
\end{eqnarray}
The real part is evaluated numerically by using the procedure
explained in Ref. \onlinecite{thomas}. We have verified that with a
cutoff $\Lambda \approx 20$  the evaluated voltage drop $\Delta V$
is independent of this cutoff.

For the reservoirs, we consider a constant density of states within a cutoff $\pm \Lambda^{\prime}$.
So, the retarded Green function for the probes can be calculated using
the Kramers-Kronig relations and gives
\begin{equation}
  g^R_j (0, \omega+\mu) = \frac{1}{2 \Lambda^{\prime} }
  \mbox{ln} | \frac{ \omega + \Lambda^{\prime} }{\omega - \Lambda^{\prime} } |
  -\frac{i\pi}{2 \Lambda^{\prime}} \Theta(\Lambda^{\prime} - |\omega|)
  \label{eq:gretreservoir}
\end{equation}




\begin{thebibliography}{11}
\bibitem{note} Although we have set $e=\hbar= v=1$ along
this work, we restore these constants along this section in order to
express the units of the different quantities here introduced in a
more transparent way.
\end{thebibliography}
\end{document}